\documentclass{iau}
\usepackage{graphicx}

\title[Estimation of the accretion mass from the statistics of neutron star masses] 
{Impact of accretion on the statistics of neutron star masses}

\author[Z. Chengg et al.]   
{Z. Cheng$^\dag$, A. Taani and Y.H. Zhao}

\affiliation{National Astronomical Observatories, Chinese Academy of
Sciences, Beijing 100012, China
\\$^\dag$email: {\tt chengzheng@bao.ac.cn}
\\[\affilskip]
}

\pubyear{2012}
\volume{290}  
\pagerange{119--126}
\jname{Feeding compact objects: Accretion on all scales}
\editors{C.M. Zhang, T. Belloni, M. M\'endez \& S.N. Zhang, eds.}
\begin{document}

\maketitle

\begin{abstract}
We have collected the parameter of 38 neutron stars (NSs)
in binary systems with spin periods and measured masses. By
adopting the Boot-strap method, we reproduced the procedure of mass
calculated for each system separately, to determine the
truly mass distribution of the NS that obtained from observation.
We also applied the Monte-Carlo simulation and introduce the
characteristic spin period 20 ms, in order to distinguish between
millisecond pulsars (MSPs) and less recycled pulsars. The mass
distributions of MSPs and the less recycled pulsars could be fitted
by a Gaussian function as $\rm 1.45\pm0.42 M_{\odot}$ and
$\rm 1.31\pm0.17 M_{\odot} ( \rm with ~ 1\sigma)$ respectively. As
such, the MSP masses are heavier than those in less
recycled systems by factor of $\rm \sim 0.13M_{\odot}$,
since the accretion effect during the recycling process.

\end{abstract}
\firstsection 
\section{Content}
\begin{figure}[h]
\centering
\includegraphics[scale=0.4]{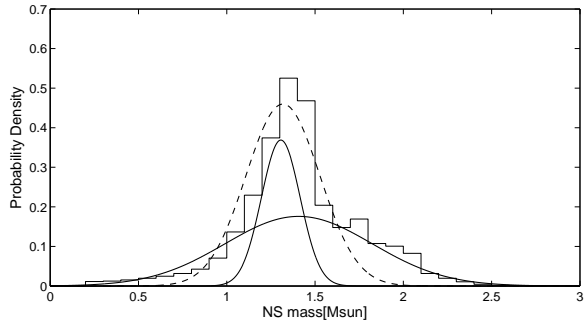}
\caption{Monte Carlo simulation of NS masses}
\label{f1}
\end{figure}

\begin{figure}[h]
\centering
\includegraphics[scale=0.5]{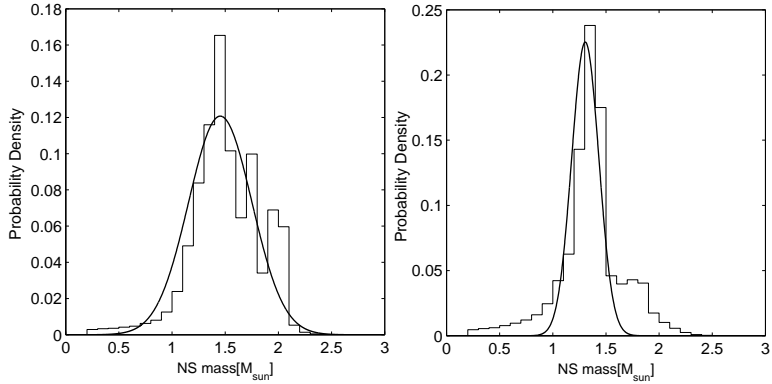}
\caption{Monte Carlo simulation of mass for MSPs (Left) and  less recycled
pulsars (Right).} \label{f2}
\end{figure}

The measured mass distribution of NSs in pulsar binary systems is
more diverse than previously thought. 
We have applied the Boot-strap method together with the
Monte-Carlo simulation to the eclipse X-ray binaries, low mass
X-ray binaries, double neutron stars (DNS) and white dwarf (WD)-NS
systems, respectively. The result of this simulation is shown in
Fig.~\ref{f1}.  Fitting a Gaussian function gives us the
mass distribution is $\rm 1.31\pm0.29M_{\odot}$ with 1$\rm \sigma$
confidence level (dash line). While the bimodal
distributions can be fitted the simulated result perfectly. This
result agrees with some theoretical predictions such as
those by Kiziltan et al. (2010) and Valentim et al. (2011). The
solid line shows the double Gaussian fitting curves ($\rm
1.30\pm0.15M_{\odot}$ and $\rm 1.41\pm0.59M_{\odot}$ with 1$\sigma$
confidence level, respectively). It should be noticed that
the bimodal distribution is coincidence with the result of Bayesian
statistical method based on the data of DNS and NS-WD systems as
showed by Kiziltan et al. (2010). This has strongly supported the
idea that the recycled MSP become more massive ($\rm
\sim0.13M_{\odot}$) than those in less recycled pulsar via
the accretion phase in recycling process.
To discriminate clearly whether this result is due to the selection effect 
or different evolution scenario, we set a characteristic spin period 
($\rm P_{spin}$ = 20 ms) which helps us to divide all the sources into two groups: MSPs with  $\rm
P_{spin} <20 ~ms$ and less recycled pulsars with $\rm P_{spin} >
20~ms$ (zhang et al. 2011). We find that the average distribution of MSPs is $\rm
1.48\pm0.53 M_{\odot}$ while in less recycled pulsar is $\rm
1.35\pm0.27 M_{\odot}$. The results are shown in Fig.~\ref{f2} which
indicate that the mass of rapidly rotating MSPs is systematic higher
than those in the less recycled ones by factor of $\rm \sim
0.13M_{\odot}$ during the
accretion phase. 
 Furthermore, the large deviation of the MSP
distribution could be reduced if we invoke the
accretion-induced collapse (AIC) in WDs, since the AIC is
expected to produce normal NSs, which in binaries can evolve into
MSPs through the usual recycling scenario.

\section{Conclusion}

The simulation results of the average masses show that the
MSPs are heavier than those in less recycled pulsars. This
result strongly support the idea that the MSPs accreted $\rm
\sim0.13M_{\odot}$ during the accretion phase in the recycling
process. Furthermore, the AIC scenario must be invoked in
order to reduce the large deviations of the MSP distribution. This
work is supported by NBRPC(2009CB824800, 2012CB821800) and the
NSFC(10773017, 11173034).

\end{document}